\documentclass[10pt,aps,pra,twocolumn,superscriptaddress,longbibliography,
amsmath,amssymb,groupaddress,
]{revtex4-2}
\usepackage[colorlinks,allcolors=blue]{hyperref}

\usepackage{float}
\usepackage{xcolor}
\usepackage{graphicx}
\usepackage{dcolumn}
\usepackage{bm}
\usepackage{bbm}
\usepackage[normalem]{ulem}

\usepackage[T1]{fontenc}  
\usepackage{ulem}

\makeatletter
\newcommand{\thickhline}{%
    \noalign {\ifnum 0=`}\fi \hrule height 1pt
    \futurelet \reserved@a \@xhline
}
\newcolumntype{"}{@{\hskip\tabcolsep\vrule width 1pt\hskip\tabcolsep}}
\makeatother

\begin{document}

\title{Robustness of the quantum Mpemba effect against state-preparation errors }

\author{Matthew Mackinnon}\thanks{Corresponding author. Email: mmackinnon01@qub.ac.uk}
\affiliation{Centre for Quantum Materials and Technologies, School of Mathematics and Physics, Queen’s University Belfast, BT7 1NN Belfast, UK}

\author{Mauro Paternostro}
\affiliation{Universit\`a degli Studi di Palermo, Dipartimento di Fisica e Chimica - Emilio Segr\`e, via Archirafi 36, I-90123 Palermo, Italy}
\affiliation{Centre for Quantum Materials and Technologies, School of Mathematics and Physics, Queen’s University Belfast, BT7 1NN Belfast, UK}

\date{\today}

\begin{abstract}
The quantum Mpemba effect (QME) is a phenomenon observed in many-body systems where initial system configurations farther from equilibrium can be observed to equilibrate faster than configurations that are closer to it. By considering noise induced error in the initial system state preparation, we analyse the robustness of various models exhibiting the QME. We demonstrate that exponentially accelerated thermalisation in open system dynamics modelled by a Gorini-Kossakowski-Sudarshan-Lindblad master equation is highly sensitive to noise induced deviations in the initial state, making this approach to accelerated thermalisation difficult to achieve. In contrast, we demonstrate that accelerated restoration of symmetry in $U(1)$ symmetric random unitary circuits via increased initial symmetry breaking is robust in the presence of state preparation error. When large errors are present in the state preparation, we show that this can in fact induce a higher rate of symmetry restoration and a stronger QME.
\end{abstract}

\maketitle


The Mpemba effect is an intriguing phenomenon first reported by Erasto Mpemba in 1969~\cite{E_B_Mpemba_1969} where, under suitable conditions, hot water can freeze faster than water that is initially cooler. The source of this phenomenon remains an open area of debate and interest. In recent years, there has been interest in identifying an analogy, and thus a quantum Mpemba effect (QME), which leverages the dynamics of both open and closed quantum systems~\cite{ares2025quantummpembaeffects, Moroder_2024, bhore2025, e27060581, PhysRevLett.133.010402, Zhang_2025}. 

Consider a dynamics leading to an equilibrium steady state $\rho_{ss}$, and let us introduce a distance $\Delta$ between two states $\rho_1(t)$ and $ \rho_2(t)$ at time $t$. One could consider, as a {\it generally distinctive} signature of the QME, the existence of a time $t_c$ such that for all $t>t_c$, $\Delta(\rho_2(t), \rho_{ss}) > \Delta(\rho_1(t), \rho_{ss})$ despite $\Delta(\rho_1(0), \rho_{ss}) > \Delta(\rho_2(0), \rho_{ss})$. 
That is, a state that is initially farther from equilibrium reaches $\rho_{ss}$ more rapidly. Such an effect could be leveraged to great advantage for quantum tasks. One example is that of performing linear algebra operations utilising thermodynamic computing~\cite{thermo_computing,aifer2024thermodynamiclinearalgebra}, where the calculation to be performed is encoded into the dynamics of a quantum system and the solution is extracted from the steady state of the system after thermalisation. For such a computational scheme, the maximum rate at which computations can be performed is limited by the rate at which the system thermalises. Leveraging the QME would facilitate greatly increased processing speed for a thermodynamic computation.

 One approach to generation of the QME, proposed in Ref.~\cite{Carollo_2021}, is through accelerated thermalisation of open quantum systems, where a system is prepared in a state that has no overlap with the slowest decaying mode of the dynamics. Such a starting state may be found to initially be further from equilibrium, but elimination of the slow decaying mode's contribution to the evolution results in an exponential speed up towards thermalisation, thus featuring QME. This has been observed experimentally in setups such as trapped ions~\cite{Zhang_2025, PhysRevLett.133.010403}, superconducting qubits~\cite{xu2025observationmodulationquantummpemba} and nuclear spin systems~\cite{chatterjee2025directexperimentalobservationquantum}. Very recently, a  non-Markovian version of the QME in open quantum system has been defined and characterised~\cite{Clark}.

An instance of QME in closed system dynamics has been predicted for a closed many-body quantum system evolving under a $U(1)$ random unitary circuit (RUC). Due to the conserved charge of such a system, the eigenstate thermalisation hypothesis (ETH)~\cite{PhysRevA.43.2046, Rigol_2008} states that the reduced state of a small subsystem will equilibriate towards the corresponding grand canonical ensemble. By considering larger breaks of symmetry in the initial state of the system, it has been shown that such subsystems will thermalise to the grand canonical ensemble more rapidly while initially being farther from equilibrium~\cite{Liu_2024, turkeshi2024quantummpembaeffectrandom}. 

From both instances described above, the crucial role of the initial preparation of the system for the observation of QME is evident. In the current noisy intermediate-scale quantum (NISQ) era of quantum computation, where error-free state preparation is difficult to achieve~\cite{Leymann_2020, Rindell_2023, Coello_P_rez_2022}, the usefulness of the QME phenomenon relies on the extent to which it is able to tolerate imperfections in the state-initialisation step. Yet, there is currently a lack of analysis in literature in this respect.

In this paper, we show that the impact of state-preparation error is highly dependent on the model and approach to accelerated thermalisation. We show that the speed-up in thermalisation presented in Ref.~\cite{Carollo_2021} reduces logarithmically as errors are introduced in the initial-state preparation, thus requiring exponentially decreasing state preparation error to observe a meaningful speed-up in thermalisation. In comparison, we show that closed system approaches utilising charge-conserving RUCs are robust in the presence of noise, with minimal changes to thermalisation speed for small amounts of noise and an increase in thermalisation speed for larger degrees of noise, in certain specific cases.

The remainder of this paper is structured as follows: Sec.~\ref{models} outlines the models under consideration, along with the state preparation techniques necessary to observe the QME. A summary of the mechanism that drives the accelerated thermalisation in each case is given. In Sec.~\ref{analysis} the procedure by which noise is introduced into the state preparation is described and the impact of noisy state preparation on each of the models is investigated. Finally, in Sec.~\ref{conc} a summary of results and comparison of the models is made.

\section{Two models for QME}\label{models}

\subsection{Accelerated thermalisation in an open quantum system}\label{open system}

We first summarise the methodology presented in Ref.~\cite{Carollo_2021} to achieve exponentially accelerated thermalisation in open quantum systems. The time evolution of an open quantum system interacting with an environment via time-independent Markovian dynamics can be modelled using a Gorini-Kossakowski-Sudarshan-Lindblad (GKSL) master equation~\cite{PRXQuantum.5.020202, Manzano_2020, 10.1093/acprof:oso/9780199213900.001.0001}. The state of the system, described by density matrix $\rho$, evolves under the Liouvillian $\mathcal{L}$ as
\begin{equation}
    \dot{\rho} = \mathcal{L}[\rho] = -i[H,\rho]+\sum_{\mu=1}^{N}\lambda_\mu(L_{\mu}\rho L_{\mu}^\dagger-\frac{1}{2}\{L_{\mu}^\dagger L_\mu, \rho\}).
    \label{lindblad map}
\end{equation}
This describes the rate of change of $\rho$ as a function of the system Hamiltonian $H$ and a set of $N$ jump operators $\{L_\mu\}$, each associated with a process occurring at a rate $\lambda_\mu$. It is convenient to recast this dynamics in vector form~\cite{Gyamfi_2020, zicari2025criticalityamplifiedquantumprobingspontaneous} by changing the density matrix into the column vector
\begin{equation}
    \label{vectorisation identity}
    \rho = \begin{pmatrix}
    a & b \\
    c & d\\
    \end{pmatrix} \rightarrow |\rho\rangle\rangle = \begin{pmatrix}
    a \\ c \\
    b \\ d\\
    \end{pmatrix},
\end{equation}
and using the operator identity 
\begin{equation}
    \label{operator vectorisation identity}
    A\rho B \rightarrow \hat{O}|\rho\rangle\rangle= (B^T\otimes A)|\rho\rangle\rangle.
\end{equation}
Using these, 
the master equation 
takes the form
\begin{equation}
\begin{aligned}
    \label{vectorised lindblad map}
    &\hat{\mathcal{L}}|\rho\rangle\rangle {=}  -i(\mathbbm{1}{\otimes} H {-} H^T {\otimes} \mathbbm{1})|\rho\rangle\rangle \\ &{+} \sum_{\mu=1}^{N}\lambda_\mu\left(L_\mu^*{\otimes} L_\mu {-} \frac{1}{2}\left[\mathbbm{1}\otimes L_\mu^\dagger L_\mu {+} (L_\mu^\dagger L_\mu)^T {\otimes} \mathbbm{1}\right]\right)|\rho\rangle\rangle.
\end{aligned}
\end{equation}
This process allows the Liouvillian superoperator $\hat{\mathcal{L}}$ to be expressed as a $d^2$-dimensional matrix with $d$ the dimension of the Hilbert space of the system. We define the left and right eigenvectors of $\hat{\mathcal{L}}$, such that $\hat{\mathcal{L}}|R_\alpha\rangle\rangle = l_\alpha|R_\alpha\rangle\rangle$ and $\langle\langle L_\alpha|\hat{\mathcal{L}} = \langle\langle L_\alpha|l_\alpha$, normalised to be biorthogonal as $\langle\langle L_\alpha|R_\beta\rangle\rangle = \delta_\alpha^\beta$. Here, we can use the fact that, in the vectorised representation, the inner product between two matrices is given by the Hilbert-Schmidt inner product 
\begin{equation}
    \label{hilbert-schmidt inner}
    \langle\langle\sigma|\rho\rangle\rangle =\text{Tr}[\sigma^\dagger\rho].
\end{equation}
As it is possible to decompose the Liouvillian as
    $\hat{\mathcal{L}} = \sum_{\alpha=1}^{d^2} l_\alpha|R_\alpha\rangle\rangle\langle\langle L_\alpha|$,
the formal solution to Eq.~\eqref{vectorised lindblad map} can be expressed as
\begin{equation}
\label{liouvillian solution}
\begin{aligned}
    |\rho(t)\rangle\rangle = e^{\hat{\mathcal{L}}t}|\rho\rangle\rangle &= \left[ \sum_{\alpha=1}^{d^2} e^{l_\alpha t}|R_\alpha\rangle\rangle\langle\langle L_\alpha|\right]|\rho\rangle\rangle\\
    &=\sum_{\alpha=1}^{d^2} e^{l_\alpha t}c_\alpha|R_\alpha\rangle\rangle,
\end{aligned}
\end{equation}
where each $|R_\alpha\rangle\rangle$ represents a mode of the system dynamics with associated rate $l_\alpha$, and we have introduced the ${\mathbb C}$-numbers $c_\alpha = \langle\langle L_\alpha|\rho\rangle\rangle$, which quantify the overlap of initial state $|\rho\rangle\rangle$ with  mode $|R_\alpha\rangle\rangle$. 
The complete positivity of the dynamics enforces that $\text{Re}(l_\alpha)\le0$. In addition, the preservation of trace requires that  at least one $l_\alpha$ must be zero. Therefore, we can order these coefficients so that $0 = \text{Re}(l_1) \ge \text{Re}(l_2)\ge\ldots\ge\text{Re}(l_{d^2})$. Clearly, $\langle\langle L_1|$ is the vectorised form of the identity matrix, and thus $\langle\langle L_1|\rho\rangle\rangle =\text{Tr}[\rho]=1$. Eq.~\eqref{liouvillian solution} can thus be simplified to
\begin{equation}
    \label{decomposition with thermal state}
     |\rho(t)\rangle\rangle=|R_1\rangle\rangle +\sum_{\alpha=2}^{d^2} e^{l_\alpha t}c_\alpha|R_\alpha\rangle\rangle.
\end{equation}
As 
the real part of all non-zero $l_\alpha$'s are negative, 
$|R_1\rangle\rangle$ clearly represents the steady state of the system, i.e. $|\rho_{ss}\rangle\rangle=|R_1\rangle\rangle$, while the coefficients $l_\alpha$ control the rate at which the system approaches equilibrium. The largest non-zero coefficient, $l_2$, will be the dominant factor in the systems approach to equilibrium. Therefore, the relaxation timescale of the system will be approximately
\begin{equation}
    \label{relaxation time scale}
    \tau_2= \frac{1}{|{\rm Re}(l_2)|}.
\end{equation}
From Eq.~\eqref{decomposition with thermal state} we can see however, that the relaxation time scale of this mode of the dynamics will not be relevant if $c_2=0$. In this case, the dominant mode of the dynamics will instead correspond to rate $l_3$ and so have a relaxation time scale of
\begin{equation}
    \label{relaxation time scale ortho}
    \tau_{3} =\frac{1}{|{\rm Re}(l_3)|}.
\end{equation}
This constitutes an exponential speed up in the thermalisation of the system. By modifying the initial state of the system with an appropriate unitary rotation $U$ such that $\langle\langle L_2|(U^*\otimes U)|\rho\rangle\rangle=0$, the timescale of thermalisation will be reduced by a factor $\propto e^{(l_3-l_2)t}$, thus signalling the occurrence of this version of the QME.

We consider the implementation of this thermalisation speed-up method in two scenarios: the Dicke model ~\cite{Kirton_2018}, as addressed in Ref.~\cite{Carollo_2021}, and the quantum harmonic oscillator. The Dicke model describes a set of $N$ two-level systems, each coupled to a lossy bosonic mode. After performing an adiabatic elimination of the latter, the GKSL equation for this model is described by a single effective jump operator, and the Hamiltonian written as
 \begin{equation}
     \label{Dicke reduced Model}
     \Tilde{H} = \Omega S_z - \frac{4\omega g^2}{4 \omega^2 + \kappa^2}\frac{S_x^2}{N}, \quad \Tilde{L}_1 = \frac{2|g|\sqrt{\kappa}}{\sqrt{4\omega^2+\kappa^2}}\frac{S_x}{\sqrt{N}}
 \end{equation}
with longitudinal field strength $\Omega$, bosonic mode frequency $\omega$, spin-boson coupling strength $g$ and loss channel strength $\kappa$. Moreover, $S_x$ and $S_z$ are the total spin operators for the system of $N$ spin-1/2 particles.

For this model, it is possible to decompose the matrix form of $\langle\langle L_2|$ as $L_2 = \sum_i\alpha_i|\varphi_i\rangle\langle\varphi_i|$ where $\langle \varphi_i|\varphi_j\rangle = \delta_i^j$. Given an initial pure state of the system $\rho = |\psi\rangle\langle\psi|$, we define an orthonormal basis $\{|\psi_i\rangle\}$ with $|\psi_1\rangle = |\psi\rangle$. As $R_1$ is a semi-positive matrix and $\text{Tr}[L_2R_1] = 0$, for non-trivial $L_2$ there must exist some $n\neq1$ such that ${\rm sign}(\alpha_1) = -{\rm sign}(\alpha_n)$. Using this, we define a rotation $U(s)=U_B(s)U_A$ with 
\begin{equation}
        \label{u2}
        U_A = \sum_i|\varphi_i\rangle\langle\psi_i|,~
        U_B(s)=\openone{-}i\sin s \boldsymbol{F}+(1{-}\cos s)\boldsymbol{F}^2
\end{equation}
and $\boldsymbol{F}= |\varphi_n\rangle\langle\varphi_1| + h.c.$
Choosing the value of the rotation parameter $s$ as $s=\text{arctan}\left|{{\alpha_1}/{\alpha_n}}\right|$ yields a rotation that, when applied to $\rho$, results in a state with overlap coefficient $c_2 = \text{tr}[L_2^\dagger U(s)\rho U(s)^\dagger] = 0$.

To compare the robustness of this approach in the context of infinite-dimensional systems, we also consider a quantum harmonic oscillator of frequency $\omega_0$ coupled to a thermal bath at temperature $T$~\cite{Isar_2003}. This can be described by a Lindblad map with two jump operators and Hamiltonian written as
\begin{equation}
    \label{SHO}
    H=\omega_0 a^\dagger a, \quad L_1=\sqrt{\gamma(\Bar{n} + 1)}a, \quad L_2=\sqrt{\gamma\Bar{n}}a^\dagger
\end{equation}
with coupling strength $\gamma$ and average number of excitations in the bath $\Bar{n} = (e^{\beta\omega_0}-1)^{-1}$, where $\beta = \frac{1}{k_bT}$ is the inverse temperature. 

Unlike the Dicke model, the spectrum of eigenvalues for this model is degenerate in the real parts, with ${\rm Re}(l_2) = {\rm Re}(l_3)$. As a result, the construction of rotation operator used for the Dicke model is not possible for this model. As both these eigenvalues are complex, however, they correspond to dynamical modes of the system that describe evolution of coherences in the density matrix. Therefore, by diagonalising the density matrix of an initial state, the overlap with both of these modes is eliminated, yielding $c_1 = c_2 = 0$, and an exponential speed-up is observed. Given the spectral  decomposition of the initial state as $\rho_0 = \sum_{i=1}^N\lambda_i|\psi_i\rangle\langle\psi_i|$, we define a diagonalising unitary operator as
\begin{equation}
    U=[|\psi_1\rangle, |\psi_2\rangle,...,|\psi_N\rangle]\in\mathbb{C}_{N,N}
\end{equation}
such that 
\begin{equation}
    \rho_D = U^\dagger\rho_0 U =\text{diag}(\lambda_1,\lambda_2,...,\lambda_N).
\end{equation}

\subsection{Accelerated symmetry restoration in a closed many-body system}\label{closed systems}

In this Section we outline the model used in Refs.~\cite{turkeshi2024quantummpembaeffectrandom,Liu_2024} to demonstrate QME in closed system dynamics. We consider a chain of $N$ spin-1/2 particles under the action of a RUC endowed with a $U(1)$ symmetry corresponding to the total charge of the system, given by
\begin{equation}
    Q = \frac{1}{2}\sum_{i=1}^N\sigma_z^i
\end{equation}
where $\sigma_z^i$ is the Pauli-Z operator of the $i$th system. The RUC is constructed from two-qubit unitary gates $U_{ij}$ acting on sites $i$ and $j$ and chosen to respect the $U(1)$ symmetry $[U_{ij}, Q] = 0$. To achieve this, $U_{ij}$ is randomly chosen such that it can be written in a block diagonal form with respect to the charge eigensectors of the system. Each block is given by a random unitary drawn from the Haar ensemble. In the computational basis of the two sites, this takes the form 
\begin{equation}
    \label{RUC operator}
    U_{ij} = \begin{pmatrix}
    u_{00}& 0 & 0 & 0 \\
    0 & u_{11} & u_{21} & 0 \\
    0 & u_{12} & u_{22} & 0 \\
    0 & 0 & 0 & u_{33} \\
    \end{pmatrix}.
\end{equation}

\begin{figure}
    \centering
    \includegraphics[width=1\linewidth]{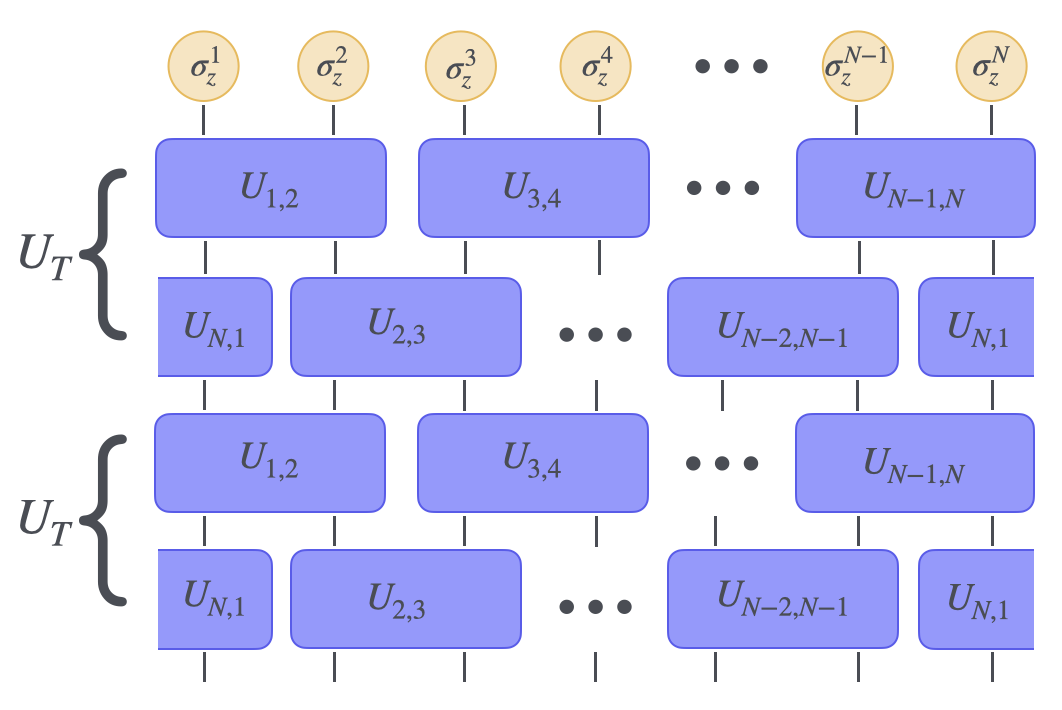}
    \caption{Schematic of two layers of a $U(1)$ symmetric RUC applied to a chain of $N$ spin-1/2 particles, with periodic boundary conditions. $U_T$ represents one complete layer of the circuit, with $U_{ij}$ representing a $U(1)$ symmetric unitary acting on spins $i$ and $j$.}
    \label{fig:ruc}
\end{figure}

Each layer of the circuit is constructed by applying a new random unitary of the form in Eq.~\eqref{RUC operator} to each adjacent pair of sites forming a composite unitary $U_T$ structured as
\begin{equation}
    U_T = \prod_i^{N-2}U_{2i-1, 2i}\prod_i^{N-2}U_{2i, 2i+1}.
\end{equation}
The structure of this circuit can be seen in Fig.~\ref{fig:ruc}, where two sequential layers with periodic boundary conditions are applied to a chain of $N$ spins.

Due to the $U(1)$ symmetry of the circuit, from ETH~\cite{PhysRevA.43.2046, Rigol_2008} the state of a small subsystem of $A\ll N$ sites will equilibrate to a grand canonical ensemble $\rho_A \propto e^{-\beta H_A-\mu Q_A}$. As a result, the state of subsystem $A$ can be expected to be symmetric in the long time limit with
\begin{equation}
    [Q,\rho_A] = 0
\end{equation}
even if the initial state of the subsystem breaks the symmetry by having a non-zero commutator. Breaking of symmetry indicates the system is out of equilibrium with restoration of subsystem symmetry representing system thermalisation. 

A common measure of distance from equilibrium in this case is the entanglement asymmetry (EA)~\cite{Ares_2023, Yamashika_2024} of the system, which measures the degree of symmetry breaking. This utilises the charge decohered state $\rho_Q$ as 
\begin{equation}
    \label{charge decoherence}
    \rho_Q = \sum_q \Pi_q \rho \Pi_q,
\end{equation}
where $\Pi_q$ is the projector onto the $q$th charge eigensector of the system. EA can then be computed as
\begin{equation}
    \label{entanglement asymmetry}
    {\rm EA}(\rho) = S(\rho_Q) - S(\rho),
\end{equation}
where $S(\rho)$ is the Von Neumann entropy of $\rho$. The entanglement asymmetry is a non-negative value and is null if and only if $\rho \equiv \rho_Q$. EA can thus be utilised as an indicator of symmetry breaking and restoration, making it a suitable measure of distance from equilibrium. A QME will be signified by an initial state of this system with a greater EA than another state, which is subsequently found to have a smaller EA at late time.

We consider the initial state of the system to be to a chain of tilted ferromagnetic spins, which reads
\begin{equation}
    \label{tilted ferromagnet}
    |\psi\rangle = \bigotimes_N|\theta\rangle\quad \text{with}\quad|\theta\rangle = e^{-i\frac{\theta}{2}\sigma_Y}|0\rangle,
\end{equation}
where $\theta$ is the tilting angle and $\sigma_Y$ the Pauli-Y operator. As the tilting of the system is increased, the conserved charge symmetry is broken due to the existence of coherences between different symmetry sectors. Increasing the tilting angle from 0 to $\pi/2$ results in an increasing degree of symmetry breaking and so the EA of the system increases. Correspondingly, an accelerated restoration of symmetry is observed for tilted ferromagnetic states that are initially more symmetry broken, and the QME is observed. Liu et al. demonstrated an increased thermalisation rate when the initial state of the system has larger overlaps $p_q = {\rm Tr}[\Pi_q\rho\Pi_q]$ with charge sectors $q$ of greater dimension $\mathcal{D}_Q$~\cite{turkeshi2024quantummpembaeffectrandom}. Charge sectors with larger $\mathcal{D}_Q$ provide more paths by which coherences can be spread across the sites of the system, resulting in a more rapid removal of coherences in the reduced state of small subsystems, and so a restoration of symmetry. Increased tilting in ferromagnetic states results in increased overlap with charge sectors with greater $\mathcal{D}_Q$, and thus accelerated thermalisation and observation of the QME.

\section{Effect of state-preparation errors on accelerated thermalisation}\label{analysis}

\subsection{Assessment of open-system model}\label{open system results}

We begin by analysing the robustness of the approach outlined in Sec.~\ref{open system}. This relies on the preparation of an initial state such that it has no overlap with the slowest decaying mode of the dynamics. This has been shown to result in an exponential acceleration of thermalisation based on the Liouvillian spectrum coefficients $l_2$ and $l_3$. 

As a measure of distance 
from equilibrium, 
we use the Hilbert-Schmidt distance~\cite{Nielsen_Chuang_2010}
\begin{equation}
\label{vectorised hilbert schmidt}
    \Delta(|\rho\rangle\rangle, |\rho_{ss}\rangle\rangle) = \langle\langle\rho -\rho_{ss}|\rho -\rho_{ss} \rangle\rangle^\frac{1}{2},
\end{equation}
where $|\rho-\rho_{ss}\rangle\rangle = |\rho\rangle\rangle - |\rho_{ss}\rangle\rangle$. This takes non-negative values which increase as the system is moved farther from equilibrium, and becomes zero only when the system has reached the steady-state $|\rho_{ss}\rangle\rangle$.

\begin{figure}
    \centering
    \includegraphics[width=1\linewidth]{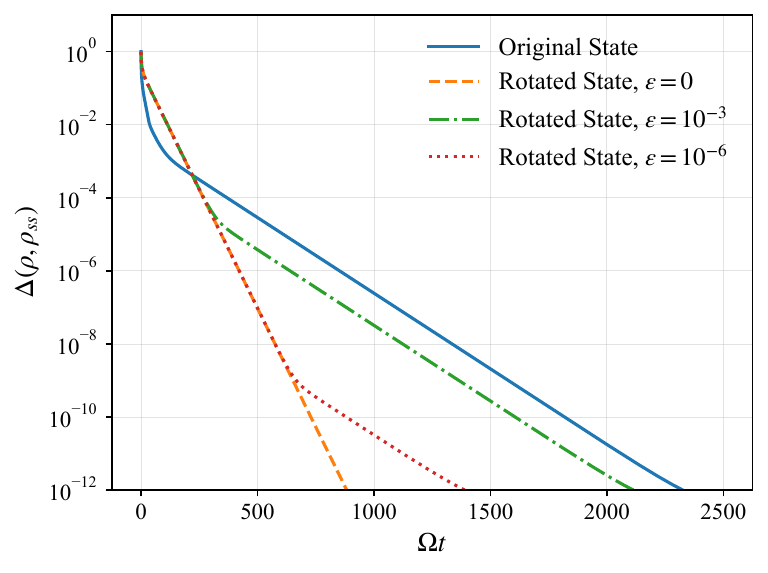}
    \caption{Hilbert-Schmidt distance $\Delta$ as a function of time $t$  for the Dicke model with $N=40$ spins. Original state is a random pure state drawn from the Fubini-Study measure presented in comparison with corresponding orthogonalised states with preparation error $\varepsilon$.}
    \label{fig:dicke_model_therm}
\end{figure}

As described in Eq.~\eqref{u2}, the preparation of a state that is orthogonal to the slowest decaying mode is generated by a unitary rotation $U(s)$ for a state specific choice of $s$. To investigate the impact of imperfect preparation of this state on thermalisation, we consider an effective value 
\begin{equation}
    \Tilde{s} = s(1+\varepsilon)
\end{equation}
with $\varepsilon$ standing for an error in the rotation angle. This results in a finite overlap with the slowest decaying mode of the dynamics that increases with $\varepsilon$. 

Fig.~\ref{fig:dicke_model_therm} shows the results reported in Ref.~\cite{Carollo_2021} for the Dicke model with $N=40$ spins and parameters $\omega = \kappa = g=1$ given in units of $\Omega$. The initial state of the system is a random pure state drawn from the Fubini-Study measure, and the same is true of all subsequent figures. Results are shown for a single initial state, as system behaviour is similar for varying states from the Fubini-Study measure. The distance from equilibrium $\Delta$ is shown as a function of time $t$, also given in units of $\Omega$, with the original state decaying at a rate proportional to $e^{l_2t}$ and the accurately orthogonalised state decaying at an accelerated rate proportional to $e^{l_3t}$. The QME effect is exhibited and characterised by the crossing occurring at $t_c\approx95$. We also include the thermalisation dynamics of orthogonalised states prepared with error $\varepsilon$ in the rotation procedure. These states initially exhibit the same exponentially increased thermalisation rate as the accurately prepared state and the presence of the QME. As error is introduced into the state preparation, the thermalisation rate exhibits a return to that of the original state mid-way through the evolution. This change occurs progressively sooner as $\varepsilon$ increases, resulting in a greatly increased total thermalisation time.

\begin{figure}
    \centering
    \includegraphics[width=1\linewidth]{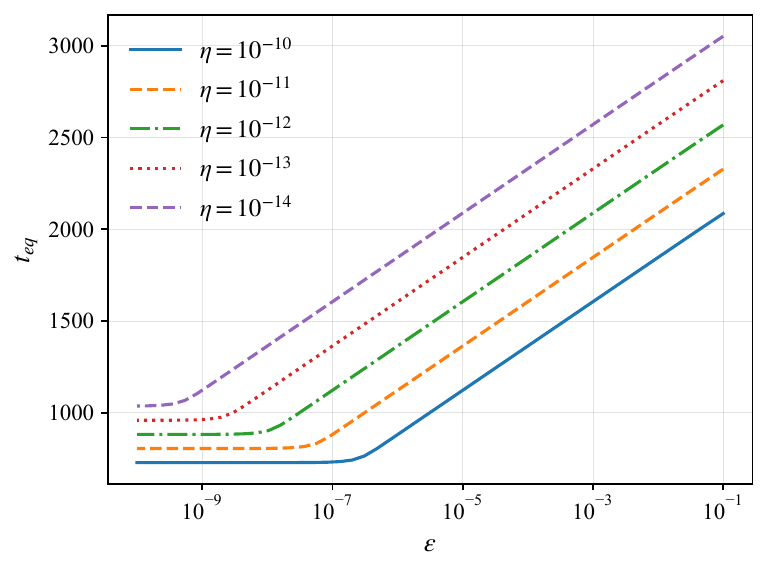}
    \caption{Time of thermalisation $t_{eq}$ as a function of state preparation error $\varepsilon$, with differing thermalisation boundaries $\eta$ for the Dicke model with $N=40$ spins.}
    \label{fig:dicke-model-therm-time-est}
\end{figure}

To investigate the relationship between error in state preparation and reduction in thermalisation speed-up, we first express $\Delta(|\rho\rangle\rangle, |\rho_{ss}\rangle\rangle)$ in terms of the overlap coefficients $c_\alpha$. The difference between $|\rho(t)\rangle\rangle$ and $|\rho_{ss}\rangle\rangle$ can be expressed as
\begin{equation}
    |\rho(t)-\rho_{ss}\rangle\rangle = \sum_{\alpha=2}^{d^2} e^{l_\alpha t}c_\alpha|R_\alpha\rangle\rangle.
\end{equation}
Using such expression, we have 
\begin{equation}
\label{Delta}
    \Delta(|\rho(t)\rangle\rangle, |\rho_{ss}\rangle\rangle) = \left[ \sum_{\alpha,\beta=2}^{N} e^{(l_\alpha^*+l_\beta) t}c_\alpha c_\beta\langle \langle R_\alpha|R_\beta\rangle\rangle\ \right]^\frac{1}{2}.
\end{equation}
As the set of states $\{|R_\alpha\rangle\rangle\}$ form an orthonormal basis, Eq.~\eqref{Delta} simplifies to
\begin{equation}
    \Delta(|\rho(t)\rangle\rangle, |\rho_{ss}\rangle\rangle) = \left[ \sum_{\alpha=2}^{N} e^{2{\rm Re}(l_\alpha) t}c_\alpha^2\right]^\frac{1}{2}.
\end{equation}
It is possible to solve for the thermalisation time $t_{eq}$ at which the state of the system reaches a given distance $\eta$ from equilibrium. By varying $\varepsilon$, the change in $t_{eq}$ can be estimated, which is shown in Fig.~\ref{fig:dicke-model-therm-time-est} for different values of $\eta$. Above a certain value of $\varepsilon$, $t_{eq}$ increases logarithmically with $\varepsilon$. As a consequence, exponentially decreasing $\varepsilon$ is required to observe a linear decrease in thermalisation time. As a result, large improvements in state preparation accuracy provide only a small decrease in thermalisation time. For a given value of $\eta$, there is a maximum value of $\varepsilon$ for which the full exponential speed-up is maintained, and this decreases linearly with $\eta$. The plateau in $t_{eq}$ for smaller values of $\varepsilon$ is due to the resulting error induced deviation in the system evolution being smaller than $\eta$ for $t<t_{eq}$ and the resulting thermalisation time matching the ideal state preparation.

\begin{figure}[b!]
    \centering
    \includegraphics[width=1\linewidth]{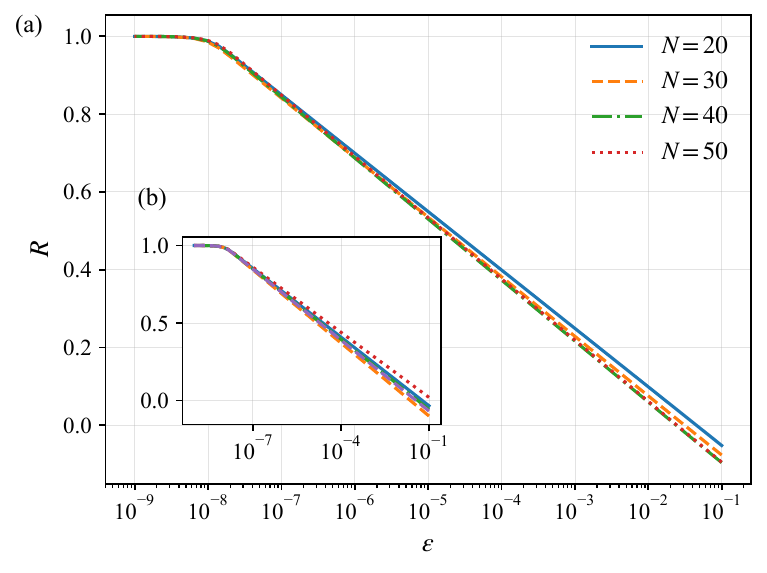}
    \caption{Relative speed-up as a function of error $\varepsilon$ for the Dicke model with (a) varying number of spins $N$ and (b) varying initial states of $N=20$ spins. Data in (a) and (b) was generated by averaging over $200$ initial states for each $N$ and using individual initial states, respectively. In all cases, $\eta=10^{-12}$ for computation of $t_{eq}$.}
    \label{fig:dicke-relative-speedup}
\end{figure}

Using the data produced for Fig.~\ref{fig:dicke-model-therm-time-est}, speed-up in thermalisation, $R$, relative to the ideal case for varying $\varepsilon$ can be expressed in terms of the unrotated and optimally rotated thermalisation times, $t_{0}$ and $t_{opt}$ respectively, as
\begin{equation}
    R = \frac{t_{0} - t_{eq}}{t_{0} - t_{opt}}.
\end{equation}
This is shown in Fig.~\ref{fig:dicke-relative-speedup} (a), where the effect of the speed-up in thermalisation is shown to decay logarithmically with $\varepsilon$. Thus, any errors in the estimation of the eigenvalues used to generate the rotation parameter $s$ will result in a significantly weaker acceleration of thermalisation. The strength of this effect can be seen to be independent of the number of spins $N$ in the model. Small variations between the gradient of lines for each size of $N$ can be observed. This is due to differences in the spectrum of coefficients $c_\alpha$ for the randomly generated states, as opposed to differences related to the dimension of the system. This is demonstrated in Fig.~\ref{fig:dicke-relative-speedup} (b), where the relative speed-up as a function of $\varepsilon$ is shown for different initial states of systems with $N=20$, showing the same variation even for states of the same number of spins. 

\begin{figure}
    \centering
    \includegraphics[width=1\linewidth]{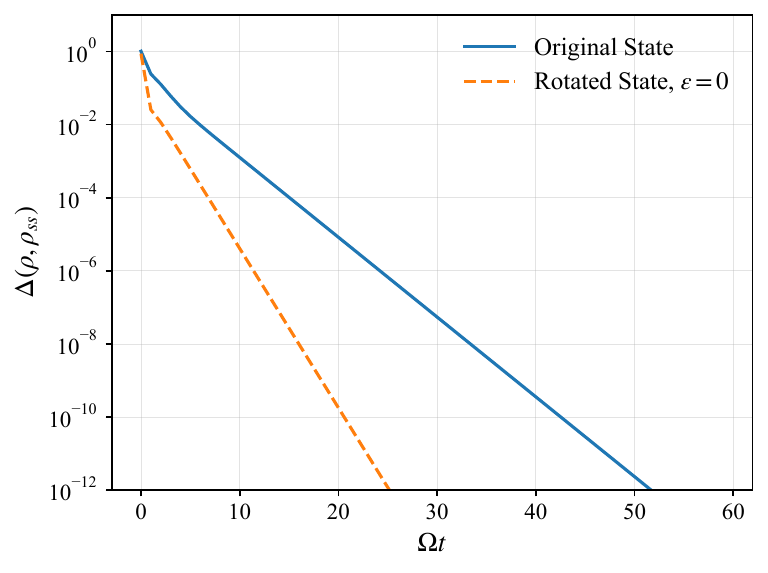}
    \caption{Hilbert-Schmidt distance $\Delta$ as a function of time $t$  for a SHO coupled to a thermal bath with occupation of up to the $N=20$th level. Original state is a random pure state presented in comparison with corresponding diagonalised state.}
    \label{fig:sho_thermalisation_dynamics}
\end{figure}

To investigate whether the impact of error on state preparation is an intrinsic feature of this approach to accelerated thermalisation or is dependent on the model used or restricted to systems of finite dimension, we also consider the model of a harmonic oscillator coupled to a finite temperature bath, as described in Sec.~\ref{open system}. Fig.~\ref{fig:sho_thermalisation_dynamics} shows $\Delta$ as a function of $t$ for a random pure state of a simple harmonic oscillator (SHO) with occupation of up to the $N=20$th level, and the associated diagonalized state. While no QME effect is observed in this setup, an exponential speed-up in thermalisation is achieved, indicated by the different gradients of the two lines.

As the unitary operator used to diagonalise the state of a SHO does not have a parameter that can be varied to introduce error into the state preparation, we consider an alternative method of modifying the rotation. To simulate state preparation error, a small perturbation can be added to the desired unitary rotation $U_T$ parameterised by the size of the perturbation $\varepsilon$. A perturbation unitary $U_P$ is drawn randomly from the Haar ensemble and a new matrix constructed as 
\begin{equation}
    U_\varepsilon = U_T + \varepsilon U_P.
\end{equation}
For small $\varepsilon$ this results in a matrix that is not necessarily unitary but is distanced from the $U_T$ based on $\varepsilon$. Using a QR decomposition~\cite{golub2013matrix}, $U_\varepsilon$ can be decomposed as a unitary $Q$ and an upper triangular matrix $R$ in the form
\begin{equation}
    U_\varepsilon = QR.
\end{equation}
This results in a unitary $Q$ that is distanced from the original matrix based on $\varepsilon$. Note, that for a given initial state, $U_P$ is taken to be constant to ensure that thermalisation time is changed only by variation of $\varepsilon$. 

\begin{figure}
    \centering
    \includegraphics[width=1\linewidth]{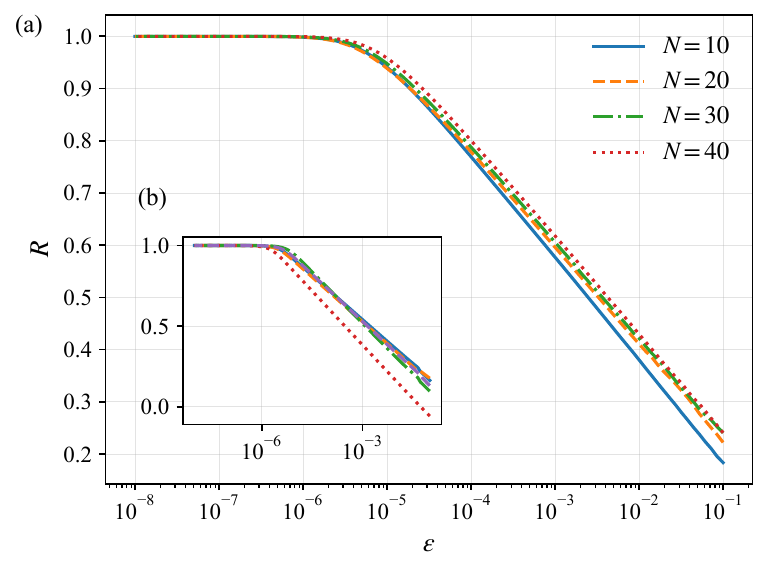}
    \caption{Relative speed-up as a function of error $\varepsilon$ for a SHO coupled to a bath with: (a) occupation of up to the $N$th level; (b) varying initial states with occupation of up to the $N=20$th level. Data in (a) and (b) was generated by averaging over $200$ initial states for each $N$ and using individual initial states, respectively. In all cases, $\eta=10^{-12}$ for computation of $t_{eq}$.}
    \label{fig:sho_thermalisation_error_impact}
\end{figure}

Utilising this process, we compute speed-up in thermalisation $R$ relative to the ideal case as a function of $\varepsilon$. This is shown in Fig.~\ref{fig:sho_thermalisation_error_impact} (a) for varying occupation of up to the $N$th energy level of the oscillator. As with the Dicke model, the relative speed-up reduces logarithmically as $\varepsilon$ increases and this effect is independent of $N$. Variations in the relative speed-up are due to the coefficients $c_\alpha$ for different initial states, and this is demonstrated in Fig.~\ref{fig:sho_thermalisation_error_impact} (b) where the same variation is observed for multiple different initial states with $N=20$ in all cases. This indicates that the susceptibility of the accelerated thermalisation to noise in the state preparation is an inherent issue with the approach that is being utilised, instead of a model-dependent issue.

\subsection{Assessment of closed-system model}\label{closed system results}

The QME has been demonstrated in $U(1)$ symmetric random unitary circuits applied to tilted ferromagnetic states as discussed in Sec.~\ref{closed systems}. We investigate the change in thermalisation speed associated with the introduction of an error $\varepsilon$ in the preparation of each spin in the tilted ferromagnetic chain. This is achieved, utilising the QR decomposition approach outlined in Sec.~\ref{open system results}, by modification of the tilting operator $U(\theta) = e^{-i\frac{\theta}{2}\sigma_Y}$ by a degree parameterised by $\varepsilon$. Each spin is modified by a different random perturbation $U_P$, allowing for randomised noise across the circuit nodes.

\begin{figure}
    \centering
    \includegraphics[width=1\linewidth]{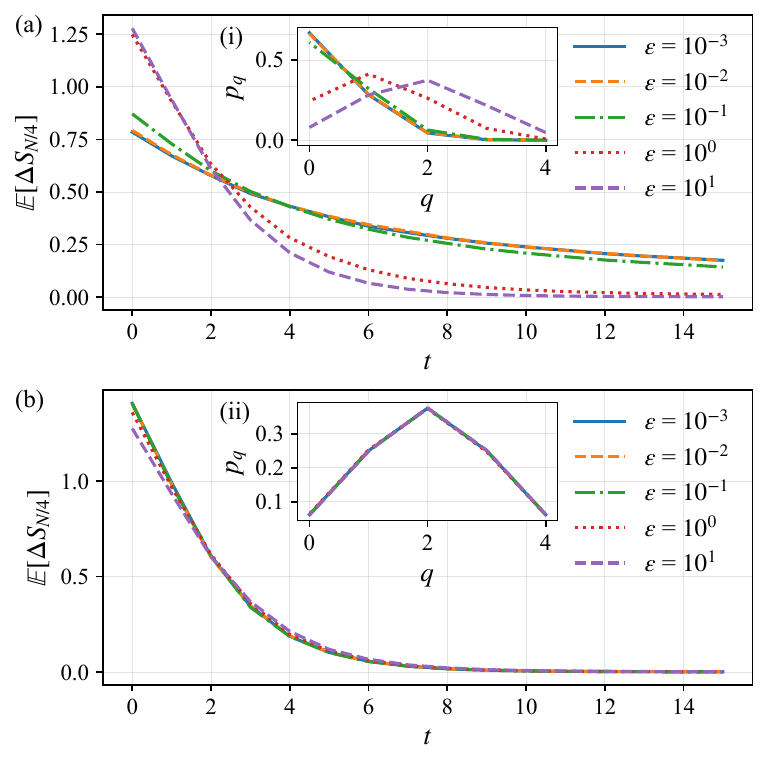}
    \caption{Analysis of random unitary circuit evolution a subsystem of $A=N/4$ sites of an $N=16$ site chain of tilted ferromagnetic states. Main plots show average entanglement asymmetry as a function of random unitary circuit depth $t$ with varying state preparation error for (a) tilting $\theta=0.2\pi$ and (b) tilting $\theta=0.5\pi$. Insets (i) and (ii) show overlap $p_q$ of initial state with each charge sector $q$ for corresponding tilting angles with varying state preparation error.}
    \label{fig:ruc_tilting_comparison}
\end{figure}

The effect of increasing $\varepsilon$ in the preparation of the initial tilted ferromagnetic state with tilting $\theta = 0.2\pi$ and $\theta=0.5\pi$ is shown in Fig.~\ref{fig:ruc_tilting_comparison} (a) and (b) respectively. In both cases, there is minimal impact on the rate of thermalisation for small values of $\varepsilon$. This remains true for all values of $\varepsilon$ in the case of $\theta=0.5\pi$. For $\theta = 0.2\pi$, large values of $\varepsilon$ result in an increased rate of thermalisation. As summarised in Sec.~\ref{closed systems}, Liu et al. conclude that the overlap $p_q$ of the initial state being skewed towards charge sectors $q$ with a larger dimension $\mathcal{D}_q$ is responsible for the observed accelerated thermalisation. Fig.~\ref{fig:ruc_tilting_comparison} (i) and (ii) show the overlap $p_q$ of the initial state with charge sector $q$, with $\mathcal{D}_q=\{1,4,6,4,1\}$ for $q=\{0, 1,2,3,4\}$. This reveals that state preparation error for tilting $\theta = 0.2\pi$ has the effect of driving the system towards a higher overlap with charge sectors of greater dimension. In the case of tilting $\theta = 0.5\pi$, all values of noise introduce no noticeable change in the overlap with different charge sectors, and hence have no impact on the rate of thermalisation.

\begin{figure}
    \centering
    \includegraphics[width=1\linewidth]{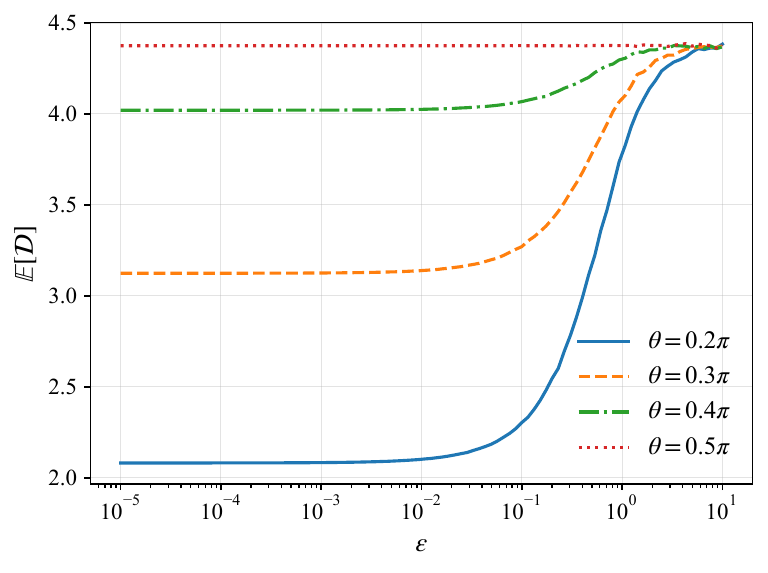}
    \caption{Weighted average charge sector dimension $\mathbb{E}[\mathcal{D}]$ as a function of $\varepsilon$ for initial tilted ferromagnetic states with different tilting angles.}
    \label{fig:overlap_vs_error}
\end{figure}

To investigate this further, we define the weighted average of charge sector dimensions as
\begin{equation}
    \mathbb{E}[\mathcal{D}] = \sum_qp_q\mathcal{D}_q
\end{equation}
to give a measure of the extent to which the initial state overlaps with charge sectors of large dimension. Fig.~\ref{fig:overlap_vs_error} shows $\mathbb{E}[\mathcal{D}]$ as a function of $\varepsilon$ for varying angles of tilting. Increasing $\varepsilon$ results in an increase in $\mathbb{E}[\mathcal{D}]$, with a rapid transition from lower to higher values between errors of the order $10^{-2}$ and $10^0$. As a consequence, noise in the state preparation has minimal impact when on the order of $10^{-2}$ and below, while above this it results in a speed up of thermalisation depending on the initial state.

Increasing $\varepsilon$ results in $U_\varepsilon$ being drawn from a distribution that becomes progressively closer to the Haar distribution. As a result, this behaviour can be explained by considering the average effect of a unitary drawn from the Haar measure~\cite{collins2006integration} applied to a density matrix of dimension $d=2$, given as
\begin{equation}
    \mathbb{E}_\text{Haar}[U\rho U^\dagger] = \frac{\mathbb{I}}{d}.
\end{equation}
The average values of $p_q$ for the state resulting from Haar averaging are the same as that of a tilted ferromagnetic state with tilting $\theta=0.5\pi$. As a result, the effect of noise on state preparation is to drive the state of the system towards a distribution of overlap values that is skewed towards charge sectors of high dimension, and correspondingly accelerates thermalisation. As a consequence, this approach to observing the QME and accelerating thermalisation is robust in the presence of noise, and can even be assisted by it.

\section{Conclusions}\label{conc}

We have outlined multiple approaches that have been previously utilised to observe the QME and facilitate accelerated thermalisation in both open and closed system dynamics. We have analysed the impact of noise in the state preparation procedure and found that the open system approach based on orthogonalisation with respect to the slow decaying modes of the dynamics is highly influenced by state preparation error. The acceleration in thermalisation is seen to logarithmically reduce as error in state preparation increases, due to an increasing overlap with the slowest decaying mode which becomes the driving factor in preventing rapid thermalisation. We have shown that this is true irrespective of system dimension, in both the case of a finite size Dicke model, and a truncated representation of an infinite dimensional SHO. As a consequence, achieving a speed-up such as this in practice would necessitate very high accuracy in the application of a many-body unitary operator that may not be feasible in the NISQ era of quantum computing.

In contrast, the accelerated thermalisation due to increased symmetry breaking in a $U(1)$ symmetric random unitary circuit has been shown to be very robust to noise in the preparation of the initial state. In the presence of large amounts of noise in the state preparation, the symmetry breaking of the initial state is found to increase on average and consequently, thermalisation is accelerated by noise in the state preparation. The driving mechanism behind this has been shown to be due to noise statistically driving the system towards states with a greater overlap with charge sectors of greater dimension.

The difference in the robustness of these approaches to facilitating accelerated thermalisation in the presence of noisy state-preparation is striking, and this highlights the necessity of error analysis to verify the robustness and feasibility of implementing such approaches. The results shown suggest that robustness or susceptibility to error is an inherent part of the approach taken to accelerated thermalisation, as opposed to being model specific. However, further analysis is required to provide a more universal characterisation of robustness of quantum Mpemba-like effects, including in the interesting instance of memory-bearing dynamics~\cite{Clark}.

\acknowledgments

MM is supported by the Department for the Economy of Northern Ireland and Equal1 Laboratories. MP acknowledges support from the European Union’s Horizon Europe EIC-Pathfinder
project QuCoM (101046973), the Department for the Economy of Northern Ireland under the US-Ireland R\&D Partnership Programme, the ``Italian National Quantum Science and Technology Institute (NQSTI)" (PE0000023) - SPOKE 2 through project ASpEQCt.

\section*{Code Availability}

The code used in this work is publicly available via Zenodo at \cite{author_code_2026}.

\bibliography{refs}

\end{document}